\documentclass[12pt]{article}

\newcommand{\be}{\begin{equation}}
\newcommand{\ee}{\end{equation}}
\newcommand{\bi}[1]{\vspace{-3mm} \bibitem{#1}}

\usepackage{epsfig,amsmath}
\usepackage{graphics}

\begin{document}

\begin{center}
{\it Chaos 16 (2006) 033108}

\vskip 5 mm
{\Large \bf Fractional Statistical Mechanics}
\vskip 5 mm

{\large \bf Vasily E. Tarasov} \\

\vskip 3mm

{\it Skobeltsyn Institute of Nuclear Physics, \\
Moscow State University, Moscow 119991, Russia } \\
{E-mail: tarasov@theory.sinp.msu.ru}
\end{center}

\vskip 3 mm

\begin{abstract}
The Liouville and first Bogoliubov hierarchy equations 
with derivatives of noninteger order are derived.
The fractional Liouville equation is obtained from the conservation 
of probability to find a system in a fractional volume element.
This equation is used to obtain Bogoliubov hierarchy and 
fractional kinetic equations with fractional derivatives.
Statistical mechanics of fractional generalization of the 
Hamiltonian systems is discussed. Liouville and Bogoliubov equations
with fractional coordinate and momenta derivatives are considered
as a basis to derive fractional kinetic equations.
The Fokker-Planck-Zaslavsky equation that has fractional 
phase-space derivatives is obtained from fractional Bogoliubov equation.
The linear fractional kinetic equation for distribution of 
the charged particles is considered. 
\end{abstract}

\vskip 3 mm

\noindent
%%%PACS: 05.20.-y; 05.20.Dd; 45.10.Hj \\
%%%Keywords: Fractional equations, Fractional derivatives, Liouville equation, 
%%%Bogoliubov equation, Fokker-Planck-Zaslavsky equation, Fractional kinetics.

%%%05.45.-a Nonlinear dynamics and nonlinear dynamical systems 
%%%05.20.-y Classical statistical mechanics
%%%05.20.Gg Classical ensemble theory
%%%05.20.Dd Kinetic theory
%%%45.10.Hj  Perturbation and fractional calculus methods

\vskip 7 mm

{\bf The theory of integrals and derivatives of noninteger order 
goes back to Leibniz, Liouville, Riemann, Grunwald, and Letnikov. 
Fractional calculus has found many
applications in recent studies in mechanics and physics.
Fractional equations, which have derivatives of noninteger order,  
are very successful in describing anomalous kinetics, transport, and chaos. 
Fractional kinetics equations usually appear from some phenomenological models. 
In this paper, we suggest fractional equations of statistical mechanics.
To obtain these equations, the conservation of probability  
to find a system in a fractional differential volume element 
of the phase-space is used. This element can be considered as 
a small part of the phase-space set with noninteger dimension.
The suggested fractional equations of statistical mechanics 
are used to derive the fractional kinetics equations. }\\

%%%%%%%%%%%%%%%%%%%%%%%%%%%%%%%%%%%%%%%%%%%%%%%%%%%%%%%%%%%%%%%%%%%%%%%%%%
\section{Introduction}

Fractional equations \cite{Podlubny,KST} contain 
derivatives of noninteger order \cite{SKM,OS}. 
Integrals and derivatives of fractional order have found many
applications in recent studies in mechanics and physics.
In a short period of time the list of such 
applications becomes long.
For example, it includes chaotic dynamics \cite{Zaslavsky1,Zaslavsky2},
mechanics of fractal media \cite{Mainardi,Media,Chaos2005},
quantum mechanics \cite{Laskin,Naber}, 
physical kinetics \cite{Zaslavsky1,Zaslavsky7,SZ,ZE,Nigmat},
plasmas physics \cite{CLZ,Plasma2005}, 
long-range dissipation \cite{GM,M,TZ2}, 
mechanics of non-Hamiltonian systems \cite{nonHam,FracHam},
theory of long-range interaction \cite{LZ,TZ3,KZT}, 
anomalous diffusion, and transport theory 
\cite{Zaslavsky1,Montr,Uch}.

Equations, which involve derivatives or integrals of noninteger order 
are very successful in describing anomalous kinetics 
\cite{Zaslavsky1,Zaslavsky2,Zaslavsky7,SZ,ZE}. 
Usually the fractional equations in dynamics or kinetics 
appear as some phenomenological models. 
In \cite{nonHam} the attempt to derive the basic 
statistical mechanics equations 
with derivatives of noninteger order have been realized.
Unfortunately, the fractional derivatives appear only 
by Fourier transform of these equations as it realized for 
the Fokker-Planck-Zaslavsky equation in \cite{Chaos2005}.

In this paper, we derive the Liouville equation with fractional 
derivatives with respect to coordinates and momenta. 
To obtain the fractional Liouville equation (FLE), we consider the 
conservation of probability in the fractional differential volume element.
This element can be considered as a small part 
of the phase-space set with noninteger dimension.
Using the FLE, we get a fractional generalization of 
the Bogoliubov hierarchy equations.
These equations can be used to derive 
fractional kinetic equations \cite{Zaslavsky1,Zaslavsky2,Zaslavsky7,SZ}.
The Vlasov equation with derivatives of noninteger order is obtained. 
The Fokker-Planck-Zaslavsky equation, which has fractional 
phase-space derivatives, is derived from the FLE.
The linear fractional kinetic equation for distribution 
of the charged particles is suggested. 

In Sec. 2, we obtain the Liouville equation with 
fractional derivatives from the conservation of probability 
in the fractional volume element of phase space. 
In Sec. 3, the first Bogoliubov hierarchy equation 
with fractional derivatives in phase space is derived.
In Sec. 4, we consider the Vlasov equation 
with fractional derivatives in phase space.
In Sec. 5, the Fokker-Planck-Zaslavsky equation, which has fractional 
derivatives with respect to coordinates and momenta is considered. 
In Sec. 6, the linear fractional kinetic equation for distribution 
of the charged particles is obtained. 
Finally, a short conclusion is given in Sec. 7.

%%%%%%%%%%%%%%%%%%%%%%%%%%%%%%%%%%%%%%%%%%%%%%%%%%%%%%%%%%%%%%%%%%%%%%%%%%
\section{Liouville equation with fractional derivatives}

A basic principle of statistical mechanics 
is the conservation of probability in the phase-space.
The Liouville equation is an expression of the  
principle in a convenient form for the analysis.
In this section, we derive the Liouville equation with 
fractional derivatives from the conservation of probability  
in a fractional volume element. 

In the phase space $R^{2n}$ with coordinates 
$(x^1,...,x^{2n})=(q_1,...,q_n,p_1,...,p_n)$, we
consider a fractional differential volume element 
\be \label{1}
d^{\alpha} V = d^{\alpha} x_1 ... \; d^{\alpha} x_{2n} .
\ee
Here, $d^{\alpha}$ is a fractional differential \cite{FDF1} 
that is defined by
\be \label{fd}
d^{\alpha} f(x)=\sum^{2n}_{k=1} D^{\alpha}_{x_k} f(x) (d x_k)^{\alpha},
\ee
where $D^{\alpha}_{x_k}$ is a fractional derivative \cite{SKM}
of order $\alpha$ with respect to $x_k$. 

The fractional derivative has different definitions \cite{OS,SKM}, 
and exploiting any of them depends on the initial (boundary) conditions, 
and the specifics of the considered physical processes.
The classical definition is the so-called Riemann-Liouville
derivative \cite{SKM}.
Due to reasons, concerning the initial and boundary conditions,
it is more convenient to use the Caputo fractional derivatives
\cite{Caputo,Podlubny}.
Its main advantage is that the initial conditions take the same
form as for integer-order differential equations.
The Caputo derivative is defined by 
\be \label{CD}
D^{\alpha}_x f(x)=\ _0^CD^{\alpha}_xf(x)=
\frac{1}{\Gamma(n-\alpha)} 
\int^x_0 \frac{f^{(n)}(z)}{(x-z)^{\alpha+1-n}} d z ,
\ee
where $n-1 < \alpha < n$, and $f^{(n)}(z)=d^n f(z)/dz^n$. 
Note that $D^{\alpha}_{x_k} 1=0$, and  $D_{x_k}x_l=0$, where $(k\not=l)$. 
Using (\ref{fd}), we obtain
\be \label{dxk}
d^{\alpha} x_k=D^{\alpha}_{x_k} x_k  (d x_k)^{\alpha} .
%%%=\frac{x^{1-\alpha}_k}{\Gamma(2-\alpha)} (d x_k)^{\alpha} .
\ee
Then
\be \label{Eq5*}
(d x_k)^{\alpha} = \left( D^{\alpha}_{x_k} x_k \right)^{-1} d^{\alpha} x_k .
\ee
From (\ref{CD}),
\be \label{CD2}
D^{\alpha}_{x_k} x^{\beta}_k=
\frac{\Gamma(\beta+1)}{\Gamma(\beta+1-\alpha)} x^{\beta-\alpha}_k  ,
\ee
where $\beta>\alpha>0$. Equations (\ref{dxk}) and (\ref{CD2}) give
\be
(d x_k)^{\alpha} =\Gamma(2-\alpha) \; x^{\alpha-1}_k d^{\alpha} x_k .
\ee

The conservation of probability for the usual phase-space volume element 
is expressed as
\be 
-d V \frac{\partial \rho (t,x)}{\partial t}=
d [\rho(t,x)\; ({\bf u},d{\bf S}) ] .
\ee
For the fractional volume element (\ref{1}), 
\be \label{duS}
-d^{\alpha} V \frac{\partial \rho(t,x)}{\partial t}=
d^{\alpha} [\rho(t,x)\; ({\bf u},d^{\alpha}{\bf S}) ] .
\ee
Here, $\rho=\rho(t,x)$ is the density of probability
to find the dynamical system in $d^{\alpha} V$, 
${\bf u}={\bf u}(t,x)$ is the velocity vector field in $R^{2n}$,
$d^{\alpha} {\bf S}$ is a surface element, and
the brackets $( \ , \ )$ is a scalar product of vectors
\be \label{uds}
{\bf u}=\sum^{2n}_{k=1} u_{k} {\bf e}_k , \quad
d^{\alpha} {\bf S}= \sum^{2n}_{k=1} d^{\alpha} S_k {\bf e}_k ,
\quad
({\bf u},d^{\alpha}{\bf S}) =\sum^{2n}_{k=1} u_k d^{\alpha} S_k ,
\ee
where ${\bf e}_k$ are the basic vectors of Cartesian coordinate system, and
\be \label{daS}
d^{\alpha}S_k =d^{\alpha} x_1 ...\; d^{\alpha} x_{k-1} d^{\alpha} x_{k+1}  
... \; d^{\alpha} x_{2n} .
\ee
The functions  $u_k=u_k(t,x)$ define $x_k$ components of ${\bf u}(t,x)$. 
In the usual case ($\alpha=1$), 
the outflow of the probability in the $x_k$ direction is 
\be \label{dru1}
d [\rho u_k ] d S_k =D_{x_k} [\rho u_k ] dx_k d S_k 
=D_{x_k} [\rho u_k ]  d V .
\ee
For $\alpha \not=1$, 
\[
d^{\alpha} [\rho u_k ] d^{\alpha} S_k 
=D^{\alpha}_{x_k} [\rho u_k ] (dx)^{\alpha} d^{\alpha} S_k . \]
Using (\ref{daS}), (\ref{1}) and (\ref{Eq5*}), we get
\be \label{dru1a}
d^{\alpha} [\rho u_k ] d^{\alpha} S_k 
= D^{\alpha}_{x_k} [\rho u_k ] \left( D^{\alpha}_{x_k} x_k \right)^{-1}
d^{\alpha} x_k d^{\alpha} S_k =
\left( D^{\alpha}_{x_k} x_k \right)^{-1}
D^{\alpha}_{x_k} [\rho u_k ]  d^{\alpha} V .
\ee
Substitution of (\ref{dru1a}) into (\ref{duS}) gives 
\be
-d^{\alpha} V \frac{\partial \rho}{\partial t}=
d^{\alpha} V \sum^{2n}_{k=1}
\left( D^{\alpha}_{x_k} x_k \right)^{-1} D^{\alpha}_{x_k} [\rho u_k ] . 
\ee

As a result, we obtain 
\be \label{cont1}
\frac{\partial \rho}{\partial t}=
-\sum^{2n}_{k=1} {\bf D}^{\alpha}_{x_k} \left( \rho u_k \right) .
\ee
where
\be
{\bf D}^{\alpha}_{x_k}=\left( D^{\alpha}_{x_k} x_k \right)^{-1}
D^{\alpha}_{x_k}=\Gamma(2-\alpha) x^{\alpha -1}_k D^{\alpha}_{x_k} .
\ee
This is the Liouville equation
with the derivatives of fractional order $\alpha$.
Equation (\ref{cont1}) describes the probability conservation for
the fractional volume element (\ref{1}) of the phase space.

For the coordinates $(q_1,...,q_n,p_1,...,p_n)$,
Eq. (\ref{cont1}) is
\be \label{Liouv1}
\frac{\partial \rho}{\partial t}
+\sum^{n}_{k=1} {\bf D}^{\alpha}_{q_k} \left( \rho V_k \right)
+\sum^{n}_{k=1} {\bf D}^{\alpha}_{p_k} \left( \rho F_k \right)=0,
\ee
where $V_k=u_k$, and $F_k=u_{k+n}$ ($k=1,...,n$).
The functions $V_k=V_k(t,q,p)$ are the components of velocity field,  
and $F_k=F_k(t,q,p)$ are the components of the force field.

In general, 
\be
D^{\alpha}_{p_k}[\rho F_k ] \not=
\rho D^{\alpha}_{p_k} F_k +F_k D^{\alpha}_{p_k} \rho .
\ee
If $F_k$ does not depend on $p_k$, and $V_k$ does not depend on $q_k$, 
then Eq. (\ref{Liouv1}) gives
\be \label{Liouv2}
\frac{\partial \rho}{\partial t}
+\sum^{n}_{k=1} \left( V_k {\bf D}^{\alpha}_{q_k} \rho 
+ F_k {\bf D}^{\alpha}_{p_k} \rho \right) =0 .
\ee
For the fractional generalization of Hamiltonian system \cite{FracHam}, 
$V_k$ and $F_k$ can be presented as 
\be \label{VF}
V_k={\bf D}^{\alpha}_{p_k} H(q,p), \quad
F_k=-{\bf D}^{\alpha}_{q_k} H(q,p) ,
\ee
where $H(q,p)$ is a fractional generalization of Hamiltonian.
Substitution of (\ref{VF}) into (\ref{Liouv2}) leads to
\be \label{Liouv3}
\frac{\partial \rho}{\partial t}
+\sum^{n}_{k=1} \left( {\bf D}^{\alpha}_{p_k} H  {\bf D}^{\alpha}_{q_k} \rho 
- {\bf D}^{\alpha}_{q_k} H {\bf D}^{\alpha}_{p_k} \rho \right)=0.
\ee

We can define 
\[
\{ A,B \}_{\alpha}=\sum^n_{k=1} 
\left( {\bf D}^{\alpha}_{q_k} A \; {\bf D}^{\alpha}_{p_k} B
- {\bf D}^{\alpha}_{q_k} B \; {\bf D}^{\alpha}_{p_k} A  \right) =
\]
\be \label{br}
=\sum^n_{k=1} 
\left( D^{\alpha}_{q_k} q_k \; D^{\alpha}_{p_k} p_k \right)^{-1}
\left( D^{\alpha}_{q_k} A \; D^{\alpha}_{p_k} B
- D^{\alpha}_{q_k} B \; D^{\alpha}_{p_k} A  \right) .
\ee
For $\alpha=1$, Eq. (\ref{br}) gives the Poisson brackets.
Note that 
\[ \{A,B\}_{\alpha}=- \{B,A\}_{\alpha}, \quad \{1,A\}_{\alpha}=0 . \]
Using (\ref{br}), we get (\ref{Liouv3}) in the form
\be
\label{Liouv5}
\frac{\partial \rho}{\partial t}+\{\rho,H\}_{\alpha}=0.
\ee
As the result, we have the Liouville equation 
for fractional generalization of Hamiltonian systems \cite{FracHam}.
For $\alpha=1$, Eq. (\ref{Liouv5}) is the usual 
Liouville equation.

%%%%%%%%%%%%%%%%%%%%%%%%%%%%%%%%%%%%%%%%%%%%%%%%%%%%%%%%%%%%%%%%%%%%%%%%%%
\section{Bogoliubov equation with fractional derivatives}

Let us consider a classical system with fixed number $N$ of
identical particles. 
Suppose that $k$th particle is described by the 
generalized coordinates $q_{ks}$ and generalized
momenta $p_{ks}$, where $s=1,...,m$.
We use the notations
${\bf q}_k=(q_{k1},...,q_{km})$ and
${\bf p}_k=(p_{k1},...,p_{km})$.
The state of this system can be  described by 
the distribution function $\rho_{N}$
in the $2mN$-dimensional phase space:
\[ \rho_{N}({\bf q},{\bf p},t)=
\rho({\bf q}_{1},{\bf p}_{1},...,{\bf q}_{N},{\bf p}_{N},t). \]
The normalization condition is
\[ \hat I[1,...,N] \rho_N({\bf q},{\bf p},t)=1, \]
where $\hat I[1,...,N]$ is the integration with respect to
${\bf q}_{1},{\bf p}_{1},...,{\bf q}_{N},{\bf p}_{N}$.

The fractional Liouville equation is
\be \label{r2}
\frac{\partial \rho_{N}}{\partial t}=-
\sum^{N}_{k=1} \Bigl(
{\bf D}^{\alpha}_{\bf q_k} ({\bf V}_k \rho_{N})+
{\bf D}^{\alpha}_{\bf p_k} ({\bf F}_k \rho_{N}) \Bigr) , \ee
where ${\bf V}_k$ is a velocity of $k$th particle,
${\bf F}_k$ is the force that acts on $k$th particle, and
\be \label{bfD}
{\bf D}^{\alpha}_{\bf q_k} {\bf V}_k=\left( D^{\alpha}_{{\bf q}_k} 
{\bf q}_k \right)^{-1} D^{\alpha}_{\bf q_k}{\bf V}_k=
\sum^m_{s=1} \left( D^{\alpha}_{q_{ks}} q_{ks} \right)^{-1} 
D^{\alpha}_{q_{ks}} V_{ks},
\ee
\be
{\bf D}^{\alpha}_{\bf p_k}{\bf F}_k
=\left( D^{\alpha}_{{\bf p}_k} {\bf p}_k \right)^{-1} 
D^{\alpha}_{\bf p_k} {\bf F}_k
=\sum^m_{s=1} \left( D^{\alpha}_{q_{ks}} p_{ks} \right)^{-1} 
D^{\alpha}_{p_{ks}} F_{ks} .
\ee
The one-particle reduced distribution function $\rho_1$ can be defined by 
\be \label{r1}  \rho_{1}({\bf q},{\bf p},t)=
\rho({\bf q}_{1},{\bf p}_{1},t)=
\hat I [2,...,N] \rho_{N}({\bf q},{\bf p},t), \ee
where $\hat I[2,...,N]$ is an integration with respect 
to ${\bf q}_2$, ..., ${\bf q}_N$, ${\bf p}_2$, ..., ${\bf p}_N$.
Obviously, that the function (\ref{r1}) satisfies
the normalization condition 
$\hat I[1]  \rho_{1}({\bf q},{\bf p},t)=1$.

The Bogoliubov hierarchy equations \cite{Bog3,Gur,Petrina,Mart}
describe the evolution of the reduced distribution functions, and
can be derived from the Liouville equation.
To obtain the first Bogoliubov equation with fractional 
derivatives from Eq. (\ref{r2}) we consider the
differentiation of (\ref{r1}) with respect to time 
\be \label{28}
\frac{\partial  \rho_{1}}{\partial t}=
 \frac{\partial}{\partial t} \hat I[2,...,N]  \rho_{N}=
\hat I[2,...,N] \frac{\partial  \rho_{N}}{\partial t} . 
\ee
Using (\ref{r2}) and (\ref{28}), we get
\be \label{r1i}
\frac{\partial  \rho_{1}}{\partial t}=
- \hat I[2,...,N] \sum^{N}_{k=1} \Bigl(
{\bf D}^{\alpha}_{\bf q_k} ({\bf V}_{k} \rho_{N})+
{\bf D}^{\alpha}_{\bf p_k} ({\bf F}_{k} \rho_{N})\Bigr) . 
\ee

Let us consider the integration $\hat I[{\bf q}_k]$ over
${\bf q}_{k}$ for $k$th particle term of Eq. (\ref{r1i}),
\[ \hat I[{\bf q}_k]
{\bf D}^{\alpha}_{\bf q_k} ({\bf V}_k  \rho_{N}) =
\hat I[{\bf q}_k] \left( D^{\alpha}_{{\bf q}_k} {\bf q}_k \right)^{-1}
D^{\alpha}_{\bf q_k} ({\bf V}_k  \rho_{N})  =
\Gamma(2-\alpha)\hat I[{\bf q}_k] {\bf q}^{\alpha-1}_k
D^{\alpha}_{\bf q_k} ({\bf V}_k  \rho_{N})= \]
\be \label{lim2}
=\Gamma(\alpha)\Gamma(2-\alpha)\hat I^{\alpha}[{\bf q}_k] 
D^{\alpha}_{\bf q_k} ({\bf V}_k  \rho_{N})=
\Gamma(\alpha)\Gamma(2-\alpha)
\Bigl({\bf V}_k  \rho_{N} \Bigr)^{+\infty}_{-\infty}=0 , \ee
where $\hat I^{\alpha}[{\bf q}_k]$ is a fractional integration
with respect to variables ${\bf q}_k$.
In Eq. (\ref{lim2}), 
we use that the distribution $ \rho_{N}$ in the limit
${\bf q}_k \rightarrow \pm \infty$ is equal to zero.
It follows from the normalization condition.  
If the limit is  not equal to zero, then
the integration over phase space is equal to infinity.
Similarly, we obtain
\[ \hat I[{\bf p}_{k}] 
{\bf D}^{\alpha}_{\bf p_k} \Bigl({\bf F}_{k}  \rho_{N} \Bigr) \sim
\Bigl({\bf F}_{k}  \rho_{N} \Bigr)^{+\infty}_{-\infty}=0  . \]
Then all terms in Eq. (\ref{r1i}) with $k=2,...,N$ are equal to zero. 
Therefore, Eq. (\ref{r1i}) are
\be \label{r1i2} \frac{\partial  \rho_{1}}{\partial t}=
- \hat I[2,...,N]\Bigl(
{\bf D}^{\alpha}_{\bf q_1} ({\bf V}_1  \rho_{N})
+{\bf D}^{\alpha}_{\bf p_1}({\bf F}_{1}  \rho_{N})
\Bigr) . \ee
The first term in Eq. (\ref{r1i2}) can be written as
\[ \hat I[2,...,N]
{\bf D}^{\alpha}_{\bf q_k} ({\bf V}_1  \rho_{N}) 
= {\bf D}^{\alpha}_{\bf q_1}
{\bf V}_1 \hat I[2,...,N]  \rho_{N} =
{\bf D}^{\alpha}_{\bf q_1} ({\bf V}_1  \rho_{1}).\]
For the binary interactions, 
\be \label{Fie2}
{\bf F}_{1}={\bf F}^{e}_1+\sum^{N}_{k=2} {\bf F}_{1k}, \ee
where ${\bf F}^{e}_1={\bf F}^{e}({\bf q}_{1},{\bf p}_{1},t)$ 
is the external force, and
${\bf F}_{1k}=
{\bf F}({\bf q}_{1},{\bf p}_{1},{\bf q}_{k},{\bf p}_{k},t)$
are the internal forces.
Using (\ref{Fie2}), the second term in (\ref{r1i2}) is
\[ \hat I[2,...,N] 
{\bf D}^{\alpha}_{\bf p_1} ({\bf F}_1  \rho_{N})   
=\hat I[2,...,N] \Bigl(
{\bf D}^{\alpha}_{\bf p_1} ({\bf F}^{e}_1  \rho_{N}) +
\sum^{N}_{k=2}
{\bf D}^{\alpha}_{\bf p_1} ({\bf F}_{1k}  \rho_{N} ) \Bigr) = \]
\be  \label{rr1i3} =
{\bf D}^{\alpha}_{\bf p_1} ({\bf F}^{e}_1  \rho_{1}) +
\sum^{N}_{k=2}
{\bf D}^{\alpha}_{\bf p_1} \hat I[2,...,N]
\Bigl( {\bf F}_{1k}  \rho_{N} \Bigr) . \ee
We assume that the distribution function is invariant under the
permutations of identical particles.
%%%\[ \rho_N(...,{\bf q}_{k},{\bf p}_{k},...,{\bf q}_{l},{\bf p}_{l},...,t)=
%%%\rho_N(...,{\bf q}_{l},{\bf p}_{l},...,{\bf q}_{k},{\bf p}_{k},...,t) . \]
Then $\rho_{N}$ is a symmetric function, 
and all $(N-1)$ terms in Eq. (\ref{rr1i3}) are identical:  
\be \label{rhs} \sum^{N}_{k=2}  \hat I[2,...,N] \
{\bf D}^{\alpha}_{\bf p_{1s}}
\Bigl( {\bf F}_{1k}  \rho_{N} \Bigr) 
= (N-1)  \hat I[2,...,N] \
{\bf D}^{\alpha}_{\bf p_1}
\Bigl( {\bf F}_{12}  \rho_{N} \Bigr)  . \ee
Using $\hat I[2,...,N]=\hat I[2]\hat I[3,...,N]$, we have
\be \label{Eq77} 
\hat I[2,...,N] \ {\bf D}^{\alpha}_{\bf p_1}
\Bigl( {\bf F}_{12}  \rho_{N} \Bigr) =
\hat I[2] \ {\bf D}^{\alpha}_{\bf p_1} 
\Bigl( {\bf F}_{12} \hat I[3,...,N]  \rho_{N} \Bigr) 
={\bf D}^{\alpha}_{\bf p_1}
\hat I[2] {\bf F}_{12}  \rho_{2} , \ee
where
\be \label{p2}  \rho_{2}=
\rho({\bf q}_{1},{\bf p}_{1},{\bf q}_{2},{\bf p}_{2},t)=
\hat I[3,...,N]  \rho_{N}({\bf q},{\bf p},t)  \ee
is a two-particle distribution function.

Finally, we obtain 
\be \label{er1-2} \frac{\partial  \rho_{1}}{\partial t}+
{\bf D}^{\alpha}_{\bf q_1} ({\bf V}_1  \rho_{1})+
{\bf D}^{\alpha}_{\bf p_1} ({\bf F}^{e}_1  \rho_{1}) =
I( \rho_{2}) , \ee
where
\be \label{I2} I( \rho_{2})=
-(N-1) {\bf D}^{\alpha}_{\bf p_1}
\hat I[2] {\bf F}_{12}  \rho_{2}  \ee
describes a velocity of particle 
number change in $4m$-dimensional
two-particle elementary phase volume.
This change is caused by the interactions between particles. 
Equation (\ref{er1-2}) is the fractional generalization of the 
first Bogoliubov equation. 
If $\alpha=1$, then we have the first Bogoliubov equation for
non-Hamiltonian systems \cite{AP2005-1}.

%%%%%%%%%%%%%%%%%%%%%%%%%%%%%%%%%%%%%%%%%%%%%%%%%%%%%%%%%%%%%%%%%%%%%%%%%%
\section{Vlasov equation with fractional derivatives}

Let us consider the particles as statistical independent systems.
Then
\be \label{2-12}
 \rho_2({\bf q}_{1},{\bf p}_{1},{\bf q}_{2},{\bf p}_{2},t)=
 \rho_1({\bf q}_{1},{\bf p}_{1},t)
 \rho_1({\bf q}_{2},{\bf p}_{2},t) . \ee
Substitution of (\ref{2-12}) into (\ref{I2}) gives
\be \label{V-I} 
I( \rho_{2})=-{\bf D}^{\alpha}_{\bf p_1} \rho_1 
\hat I[2] {\bf F}_{12}  \rho_1({\bf q}_{2},{\bf p}_{2},t) , 
\ee
where $\rho_1=\rho_1({\bf q}_{1},{\bf p}_{1},t)$.

Let us define
\[ {\bf F}^{eff} ({\bf q}_{1},{\bf p}_{1},t)=
\hat I[2] {\bf F}_{12}  \rho_{1}({\bf q}_{2},{\bf p}_{2},t) . \]
Then, 
\be \label{Ir2} I( \rho_{2})=
- {\bf D}^{\alpha}_{\bf p_1} ( \rho_{1} {\bf F}^{eff}) . \ee
Substituting of (\ref{Ir2}) into (\ref{er1-2}), we obtain
\be \label{p1-1} \frac{\partial  \rho_{1}}{\partial t}+
{\bf D}^{\alpha}_{\bf q_1} ({\bf V}_1  \rho_{1})
+{\bf D}^{\alpha}_{\bf p_1} \Bigl(
({\bf F}^{e}_1+(N-1){\bf F}^{eff}) \rho_{1} \Bigr)=0  \ee
that is a closed equation for the one-particle
distribution function with the external force ${\bf F}^{e}_1$
and the effective force ${\bf F}^{eff}$.
Equation (\ref{p1-1}) is the fractional generalization 
of the Vlasov equation \cite{Vlasov1,Vlasov2}
that has phase-space derivatives of noninteger order.
For $\alpha=1$, we get the Vlasov equation for 
the non-Hamiltonian systems \cite{AP2005-1}.

%%%%%%%%%%%%%%%%%%%%%%%%%%%%%%%%%%%%%%%%%%%%%%%%%%%%%%%%%%%%%%%%%%%%%%%%%%
\section{Fokker-Planck-Zaslavsky equation for phase-space}

The Fokker-Planck equations with fractional coordinate derivatives
have been suggested by Zaslavsky \cite{Zaslavsky7}
to describe chaotic dynamics.
It is known that Fokker-Planck equation can be derived
from the Liouville equation \cite{Is,RL,F}.
In this section, we obtain Fokker-Planck-Zaslavsky equation 
that has fractional derivatives in phase space.

Let us consider a system of $N$ identical particles
and the Brownian particle that is described by the
distribution function 
\[ \rho_{N+1}=\rho_{N+1}({\bf q},{\bf p},Q,P,t), \]
where
\[ {\bf q}=({\bf q}_1,...,{\bf q}_N), \quad {\bf q}_k=(q_{k1},...,q_{km}) , \]
\[ {\bf p}=({\bf p}_1,...,{\bf p}_N), \quad {\bf p}_k=(p_{k1},...,p_{km})  \]
are the coordinates and momenta of the particles;
$Q=(Q_s)$ and $P=(P_s)$ ($s=1,...,m$) 
are Brownian particle coordinates and momenta.
The normalization condition is
\be \label{nc} \hat I[1,...,N,N+1]  \rho_{N+1}=1 . \ee
The distribution function for the Brownian particle
is defined by
\be \label{rB} \rho_{B}(Q,P,t)=
\hat I[1,...,N] \rho_{N+1}({\bf q},{\bf p},Q,P,t) . \ee
The Liouville equation for $ \rho_{N+1}$ is
\be \frac{\partial  \rho_{N+1}}{\partial t}-
i(L_N+L_B) \rho_{N+1}=0 ,\ee
where 
\be -iL_N\rho = \sum^{N,m}_{k,s} \Bigl(
{\bf D}^{\alpha}_{q_{ks}} (G^k_s \rho) +
{\bf D}^{\alpha}_{p_{ks}} (F^k_s \rho)  \Bigr) , \ee
\be -iL_B\rho = \sum^{N,m}_{k,s} \Bigl(
{\bf D}^{\alpha}_{Q_{s}} (g_s \rho) +
{\bf D}^{\alpha}_{P_{s}} (f_s \rho)   \Bigr) . \ee
Here, $L_N$ and $L_B$ are Liouville operators with fractional derivatives,
and 
\[ {\bf D}^{\alpha}_{A}B =\left( D^{\alpha}_{A} A \right)^{-1} 
D^{\alpha}_{A} B . \]

The functions $G^k_s$ and $F^k_s$ are defined 
by the equations of motion for particle, 
\be \label{HE1b}
\frac{d q_{ks}}{dt}=G^k_s({\bf q},{\bf p}), \quad
\frac{d p_{ks}}{dt}=F^k_s({\bf q},{\bf p},Q,P) , \quad k=1,...,N. \ee
The Hamilton equations for the Brownian particle 
\be \label{HE2b} 
\frac{d Q_{s}}{dt}=g_s(Q,P), \quad
\frac{d P_{s}}{dt}=f_s({\bf q},{\bf p},Q,P) . \ee
define $g_s$ and $f_s$.

Let us use the boundary condition in the form 
\be \label{bound2} \lim_{t \rightarrow - \infty}
\rho_{N+1} ({\bf q},{\bf p},Q,P,t) =
\rho_N({\bf q},{\bf p},Q,T) \rho_B(Q,P,t) , \ee
where 
\be \label{Gibbs} \rho_N({\bf q},{\bf p},Q,T)=
\exp \, \beta ({\cal F}-H({\bf q},{\bf p},Q))  \ee
is the canonical Gibbs distribution for
\be H ({\bf q},{\bf p},Q)=
H_N({\bf q},{\bf p})+\sum^N_{k=1} U_B({\bf q}_k,Q) . \ee
Here, $H_N$ is a Hamiltonian of an $n$-particle system, and $U_B$
is an energy of interaction between particles and Brownian particle.
If we suppose 
\be \label{Gg} G^k_s=p_{ks}/m , \quad g_s=P_s/M , \ee
then
\be  H_N({\bf q},{\bf p})=\sum^{N,m}_{k,s} \frac{p^2}{2m}+
\sum_{k<l}U({\bf q}_k,{\bf q}_l) . \ee

The boundary condition (\ref{bound2}) can be realized \cite{Zub} by the
infinitesimal source term in the Liouville equation:
\be \label{epsl}
\frac{\partial  \rho_{N+1}}{\partial t}-i(L_N+L_B) \rho_{N+1}=
- \varepsilon ( \rho_{N+1}- \rho_N  \rho_B) . \ee
Integrating (\ref{epsl}) by $\hat I[1,...,N]$, we obtain 
\be \label{rhoB}
\frac{\partial  \rho_B}{\partial t}
+\sum^{m}_{s=1} {\bf D}^{\alpha}_{Q_{s}} (g_s \rho_B)
+\hat I[1,...,N] \sum^m_{s=1}
{\bf D}^{\alpha}_{P_{s}} (f_s \rho_{N+1}) =0 , \ee
which is the Liouville equation for reduced distribution function
of the Brownian particle. 

The formal solution of Eq. (\ref{epsl}) has the form
\be \label{58} \rho_{N+1}(t)= \rho_B(t)  \rho_N
- \int^0_{-\infty} d \tau \ e^{\varepsilon \tau}
e^{-i \tau (L_N+L_B)} 
\Bigl( \frac{\partial}{\partial \tau}-i(L_N+L_B) \Bigr)
 \rho_B (t+\tau)  \rho_N . \ee
Substituting (\ref{58}) into (\ref{rhoB}), we get
\[ \frac{\partial  \rho_B}{\partial t}
+\sum^{m}_{s=1} {\bf D}^{\alpha}_{Q_{s}} (g_s  \rho_B)
+\sum^m_{s=1} {\bf D}^{\alpha}_{P_{s}}
\rho_B \hat I [1,...,N] (f_s  \rho_N)- \]
\be \label{rhoB2} 
-\hat I^{\alpha}[1,...,N]\sum^m_{s=1}
{\bf D}^{\alpha}_{P_{s}} 
\int^0_{-\infty} d \tau \ e^{\varepsilon \tau}
e^{-i \tau (L_N+L_B)} 
\Bigl( \frac{\partial}{\partial \tau}-i(L_N+L_B) \Bigr)
\rho_B (t+\tau)  \rho_N=0 . \ee
The expression $\hat I [1,..,N] f_s  \rho_N $
can be considered as average value of the force $f_s$.
For the canonical Gibbs distribution (\ref{Gibbs}) it
is equal to zero. Using
\be  
{\bf D}^{\alpha}_{Q_{s}} \rho_N=
\frac{1}{kT} f^{(p)}_s \rho_N , \ee
where $f^{(p)}_s$ is a fractional potential force \cite{nonHam}:
\be \label{fp}
f^{(p)}_s=- {\bf D}^{\alpha}_{Q_{s}} U_B , \ee
we have
\[ -iL_B  \rho_{N+1}=
\Bigl( \frac{P_sf^{(p)}_s}{MkT} \rho_B +
{\bf D}^{\alpha}_{Q_{s}} (g_s \rho_B)+
{\bf D}^{\alpha}_{P_{s}} (f_s \rho_B) \Bigr) \rho_N  . \]

It can be proved by integration that the term
\be \frac{\partial  \rho_B}{\partial t}+
{\bf D}^{\alpha}_{Q_{s}} (g_s  \rho_B) \ee
in Eq. (\ref{rhoB2}) does not contribute.
Then (\ref{rhoB2}) gives
\[
\frac{\partial  \rho_B}{\partial t}
+\sum^{m}_{s=1} {\bf D}^{\alpha}_{Q_{s}} (g_s  \rho_B)
+\sum^m_{s=1} {\bf D}^{\alpha}_{P_{s}} 
\hat I[1,...,N]  
\int^0_{-\infty} d \tau \ e^{\varepsilon \tau}
 f_s e^{-i \tau (L_N+L_B)}  \rho_N  \cdot \]
\be \label{rhoB3} 
\cdot \Bigl( {\bf D}^{\alpha}_{P_{s'}} (f_{s'}\rho_B(t+\tau))
+\frac{\beta f_{s'} P_{s'}}{M} \rho_B (t+\tau) \Bigr)=0 . \ee
Equation (\ref{rhoB3}) is a closed integro-differential equation for the
distribution function $ \rho_B$.
Note that $f_s$ can be presented as
\[ f_s=f^{(p)}_s+f^{(n)}_s ,\]
where $f^{(p)}_s$ is a potential force (\ref{fp}), and
$f^{(n)}_s$ is a non-potential force that acts on the Brownian particle.
For the equilibrium approximation
$P \sim (MkT)^{1/2}$, $iL_B \sim M^{-1/2}$ and
$iL_N \sim m^{-1/2}$. If $M>>m$, we can use perturbation theory.

Using the approximation $\rho_B(t+\tau)=\rho_B(t)$
for Eq. (\ref{rhoB3}), we obtain 
%%$m<<M$ ($L_B<<L_N$)
\be \label{rhoB4}
\frac{\partial  \rho_B}{\partial t}
+\sum^{m}_{s=1} {\bf D}^{\alpha}_{Q_{s}}(g_s \rho_B)
+\sum^m_{s=1} {\bf D}^{\alpha}_{P_{s}} 
\Bigl( \frac{M}{\beta}
{\bf D}^{\alpha}_{P_{s'}} ( \gamma^1_{ss'}  \rho_B(t))
+ \gamma^2_{ss'} P_{s'}  \rho_B (t) \Bigr)=0 , \ee
where
\be \label{gamma1}
\gamma^1_{ss'}=\beta M \hat I^{\alpha}[1,...,N]
\int^0_{-\infty} d \tau \ e^{\varepsilon \tau}
f_s e^{-i \tau L_N} f_{s' } \rho_N , \ee
\be \label{gamma2}
\gamma^2_{ss'}=\beta M \hat I^{\alpha}[1,...,N]
\int^0_{-\infty} d \tau \ e^{\varepsilon \tau}
f_s e^{-i \tau L_N} f^{(p)}_{s' }  \rho_N . \ee
If $f_s=f^{(p)}_s$, then $\gamma^1_{ss'}=\gamma^2_{ss'}$.
As a result, we derive the Fokker-Planck-Zaslavsky 
equation \cite{Zaslavsky7,Chaos2005} for the phase space.

%%%%%%%%%%%%%%%%%%%%%%%%%%%%%%%%%%%%%%%%%%%%%%%%%%%%%%%%%%%%%%%%%%%%%%%%%%
\section{Linear fractional kinetic equation}

Let us consider Eq. (\ref{er1-2})
with $I(\rho_2)=0$, ${\bf V}_1={\bf p}/m={\bf v}$, and 
${\bf F}^e=e{\bf E}$, ${\bf B}=0$. Then
\be \label{fke1} \frac{\partial  \rho_1}{\partial t}+
({\bf v}, {\bf D}^{\alpha}_{\bf q} \rho_1) +
e({\bf E},  {\bf D}^{\alpha}_{\bf p} \rho_1) =0, \ee
where 
\be
({\bf v}, {\bf D}^{\alpha}_{\bf q} \rho) =
\sum^m_{s=1} (v_s, {\bf D}^{\alpha}_{{\bf q}_s} \rho) .
\ee

If we take into account the magnetic field (${\bf B}\not=0$), 
then we must use the fractional generalization of Leibnitz rules
\be
{\bf D}^{\alpha}_{\bf p} (fg)=\sum^{\infty}_{s=0} 
\frac{\Gamma(\alpha+1)}{\Gamma(s+1) \Gamma(\alpha-s+1)}
({\bf D}^{\alpha-s}_{\bf p} f) {\bf D}^s_{\bf p} g ,
\ee
where $s$ are integer numbers. 
In this case, Eq. (\ref{fke1}) has the addition term 
\[ \frac{e}{mc}
{\bf D}^{\alpha}_{\bf p} \left( [{\bf p}, {\bf B}] \rho_1 \right)=
\frac{e}{mc} \sum_{k l m} 
{\bf D}^{\alpha}_{p_k} \left( \varepsilon_{klm} p_l B_m \rho_1 \right)
=\frac{e}{mc} \sum_{k l m} \varepsilon_{klm} B_m 
{\bf D}^{\alpha}_{p_k} \left( p_l \rho_1 \right)= \]
\[ =\frac{e}{mc} \sum_{k l m} \varepsilon_{klm} B_m 
\sum^1_{i=0}\frac{\Gamma(\alpha+1)}{\Gamma(i+1) \Gamma(\alpha-i+1)}
[{\bf D}^{\alpha-i}_{p_k} \rho_1] \delta_{kl} p^i_l= \]
\[ =\frac{e}{mc} \sum_{k l m} \varepsilon_{klm} B_m  \left(
[{\bf D}^{\alpha}_{p_k} \rho_1] p_l+
\alpha [{\bf D}^{\alpha-1}_{p_k} \rho] \delta_{kl} \right)= \]
\be
\frac{e}{mc} \sum_{k l m} \varepsilon_{klm} B_m p_l
[{\bf D}^{\alpha}_{p_k} \rho_1]=
\frac{e}{mc} \left( ({\bf D}^{\alpha}_{p_k} \rho_1),
[{\bf p}, {\bf B}] \right)  .
\ee

Let us consider the perturbation \cite{Eck,KT} 
of the distribution function in the form
\be \label{pert}
\rho_1=\tilde \rho_1 +\delta \rho_1(t,q,p) ,
\ee
where $\tilde \rho_1$ is a homogeneous stationary density of probability
that satisfies Eq. (\ref{fke1})  for ${\bf E}=0$.
Substituting (\ref{pert}) into Eq. (\ref{fke1}), we get 
\be \label{fke2} \frac{\partial \delta \rho_1}{\partial t}+
({\bf v}, {\bf D}^{\alpha}_{\bf q} \delta \rho_1) +
e({\bf E} , {\bf D}^{\alpha}_{\bf p} \tilde \rho_1 )=0. \ee
Equation (\ref{fke2}) is linear fractional kinetic equation for 
the first perturbation $\delta \rho_1$ of the distribution function.
Solutions of fractional linear kinetic equations (\ref{fke2})
are considered in Ref. \cite{SZ}.
For ${\bf E}=0$, the function $\delta \rho_1$ 
is described by the function
\be \label{HH11}
(g_st)^{-1/\alpha} L_{\alpha} \left[ q_s (g_st)^{-1/\alpha} \right],
\ee 
where $g_s=v_s (D^{\alpha}_{q_s} q_s)^{-1}$, and
\be \label{La}
L_{\alpha}[x]=\frac{1}{2\pi} \int^{+\infty}_{-\infty} dk \  
e^{-ikx} e^{-a|k|^{\alpha}} 
\ee
is the Levy stable p.d.f. \cite{Feller}.
The examples of $L_{\alpha}[x]$ 
%%%for $\alpha=2.0$, $\alpha=1.6$, and $\alpha=1.0$ 
are shown in Fig. 1.

For $\alpha=1$, the function (\ref{La}) gives
the Cauchy distribution 
\be  \label{HHH1}
L_1 [x]=\frac{1}{\pi} \frac{1}{x^2+1} ,
\ee
and (\ref{HH11}) is
\be \label{HHH2}
\frac{1}{\pi} \frac{ (g_s t)^{-1}}{q^2_s (g_st)^{-2} +1}  .
\ee 
For $\alpha=2$, Eq. (\ref{La}) gives the Gauss distribution: 
\be  \label{HHH3}
L_2 [x]= \frac{1}{2\sqrt{\pi}} e^{-x^2/4} ,
\ee
and the function (\ref{HH11}) is
\be \label{HHH4}
(g_st)^{-1/2} \frac{1}{2\sqrt{\pi}} e^{-q^2_s /(4g_s t) } .
\ee 

%%%%%%%%%%%%%%%%%%%%%%%%%%%%%%%%%%%%%%%%%%%%%%%%%%%%%%%%%%%%%%%%%%%%%%%%%%

\begin{figure}
\centering
\rotatebox{270}{\includegraphics[width=11 cm,height=15 cm]{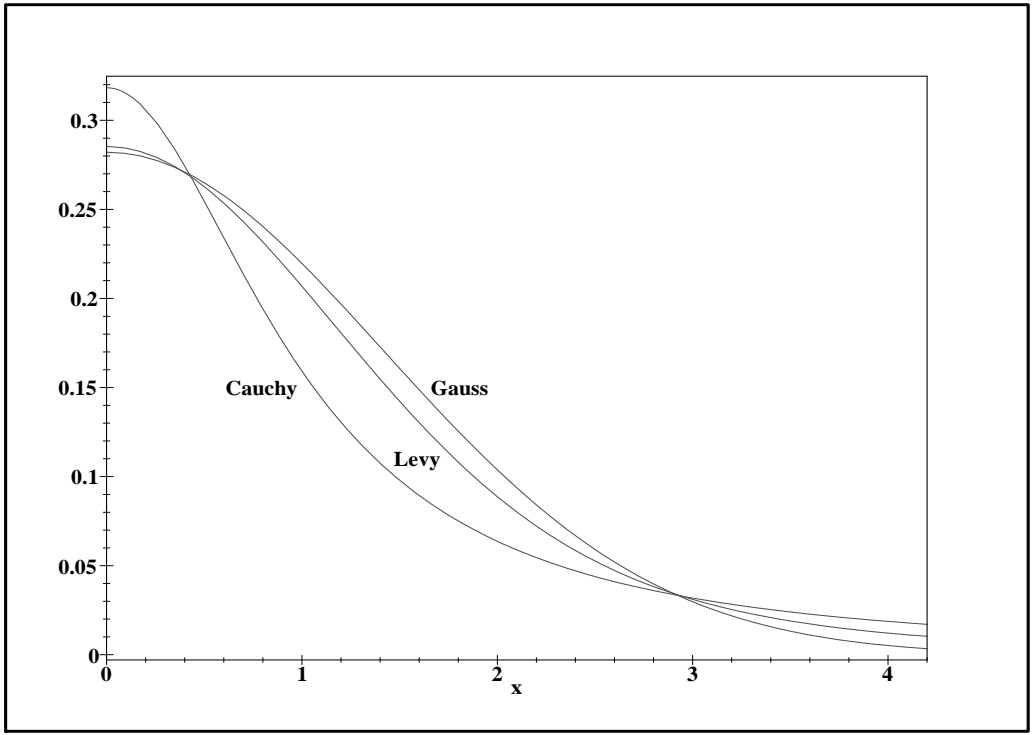}}
\caption{Gauss p.d.f. ($\alpha=2$), 
Levy p.d.f. ($\alpha=1.6$), and Cauchy p.d.f. ($\alpha=1.0$).  
Levy for $\alpha=1.6$ lies between Cauchy and Gauss p.d.f.
In the asymptotic $x \rightarrow \infty$ and $x>3$ on the plot, 
the upper curve is Cauchy p.d.f., and the lower curve is the Gauss p.d.f.}
\end{figure}

%%%%%%%%%%%%%%%%%%%%%%%%%%%%%%%%%%%%%%%%%%%%%%%%%%%%%%%%%%%%%%%%%%%%%%%%%%

For $1< \alpha \le 2$, the function $L_{\alpha}[x]$ can be presented  
as the expansion
\be
L_{\alpha}[x]=-\frac{1}{\pi x} \sum^{\infty}_{n=1} 
(-x)^n \frac{\Gamma(1+n/\alpha)}{n!} \sin (n \pi/2) .
\ee
The asymptotic ($x \rightarrow \infty$, $1<\alpha<2$) is given by
\be
L_{\alpha}[x] \sim -\frac{1}{\pi x} \sum^{\infty}_{n=1} 
(-1)^n x^{-n \alpha} \frac{\Gamma(1+n \alpha)}{n!} \sin (n \pi/2)  .
\ee
As the result, the asymptotic of the solution, 
exhibits the power-like tails for $x \rightarrow \infty$.
This tails is the important property of solutions of
equations with the noninteger derivative.

%%%%%%%%%%%%%%%%%%%%%%%%%%%%%%%%%%%%%%%%%%%%%%%%%%%%%%%%%%%%%%%%%%%%%%%%%%
\section{Conclusion}

In this paper, we consider equations with derivatives 
of noninteger order that can be used in statistical mechanics 
and physical kinetics.
We derive the Liouville, Bogoliubov, Vlasov 
and Fokker-Planck equations with fractional derivatives 
with respect to coordinates and momenta. 
To derive the fractional Liouville equation (FLE), 
we consider the conservation of probability 
in the fractional differential volume element.
This element can be considered as a small part of 
the phase-space set with noninteger dimension.
Using the FLE, we obtain
a fractional generalization of the Bogoliubov hierarchy equations.
These equations describe the evolution 
of the reduced density of probability in the fractional 
phase-space volume element.
Fractional Bogoliubov equations can be used to derive 
fractional kinetic equations \cite{Zaslavsky1,Zaslavsky7,SZ}.
In this paper, we obtain Fokker-Planck-Zaslavsky equation,  
fractional Vlasov and linear kinetic equations. 

The fractional kinetics is related to the equations 
that contains derivatives of noninteger order.
These equations appear in the description of chaotic 
dynamics, and the fractal media.
The fractional derivatives can be connected with long-range 
power-law interaction of the systems \cite{LZ,TZ3,KZT}. 
For noninteger derivatives with respect to coordinates, 
we have the power-like tails as the important property 
of the solutions of the fractional equations.

%\newpage
%%%%%%%%%%%%%%%%%%%%%%%%%%%%%%%%%%%%%%%%%%%%%%%%%%%%%%%%%%%%%%%%%%%%%%%%%%

\end{document}